\documentclass[]{spie}  %>>> use for US letter paper
\usepackage{amsmath,amsfonts,amssymb}
\usepackage{graphicx}
\usepackage{comment}
\usepackage{tikz}
\usepackage{pgf}
\usepackage{pgfplots}
\pgfplotsset{compat=1.18}
\usetikzlibrary{positioning}
\usetikzlibrary{arrows.meta, positioning, shapes.geometric, calc, decorations.pathreplacing}
\usepackage{float}

\title{Operational capabilities and on-sky performance of SAMOS at the completion of science commissioning}

\author[a,b]{Massimo Robberto}
\author[b]{Stephen A. Smee}
\author[b]{Robert H. Barkhouser}
\author[b]{Stephen C. Hope}
\author[c]{John J. Piotrowski}
\author[b]{Dana Koeppe}
\author[a]{Mario Gennaro}
\author[d]{Zoran Ninkov}
\author[e]{Megan E. Donahue}
\author[f]{Andrei Tokovinin}
\author[b]{Randolph P. Hammond}
\author[b]{Albert J. Harding}

\affil[a]{Space Telescope Science Institute, 3700 San Martin Drive, Baltimore, MD 21218, USA}
\affil[b]{Department of Physics and Astronomy, Johns Hopkins University, Baltimore, MD 21218, USA}
\affil[c]{The Observatories of the Carnegie Institution for Science, Pasadena, CA, USA}
\affil[d]{Rochester Institute of Technology, Rochester, NY, USA}
\affil[e]{Michigan State University, East Lansing, MI, USA}
\affil[f]{Cerro Tololo Inter-American Observatory, NSF NOIRLab, La Serena, Chile}

\authorinfo{
Further author information: send correspondence to M.R.\\
M.R.: E-mail: robberto@stsci.edu; Telephone: +1 410 338 4382, orcid: 0000-0002-9573-3199
}

\makeatletter
\let\addcontentsline\@gobblethree
\let\@writefile\@gobbletwo
\makeatother

\begin{document} 
\maketitle

\begin{abstract}
We present the operational capabilities and performance of the SOAR Adaptive Module Optical Spectrograph (SAMOS) at the completion of its science commissioning phase. SAMOS is a Digital Micromirror Device (DMD)-based multi-slit spectrograph and imager deployed behind the SOAR Adaptive Module (SAM) ground-layer adaptive optics system. SAMOS relays the full AO-corrected $3'\times3'$ field of view onto the central 1K×1K region of a large-format DMD. Each DMD mirror subtends $0.17"\times0.17"$
 on the sky and directs light to a spectroscopic or a parallel imaging channel, allowing the generation of programmable slit-mask patterns with near-instantaneous reconfiguration. The spectrograph employs two low-resolution gratings covering the 4000–10000\,\AA\, wavelength range at resolving powers of R $\sim$2500. High-resolution spectroscopy ($R\sim10\,000$) is available in two wavelength ranges, 4500--5150 \AA \ and 6000--7000 \AA, for a 0.33'' slit width.
 A fifth grating, currently under contract, will provide spectroscopy R$\sim$ 10,000 in the region of the Ca II radial velocity spectrometer of Gaia (8500–8750 \AA). We summarize the operational workflow established during commissioning, including astrometric registration, DMD mask generation, simultaneous imaging and spectroscopy, and automated data reduction. Performance metrics derived from science-verification observations demonstrate the instrument's multiplexing capability, wavelength-calibration accuracy, and end-to-end photometric calibration. These results establish SAMOS as a unique facility instrument combining adaptive-optics-assisted imaging, programmable multi-object spectroscopy, and rapid observational flexibility.
\end{abstract}

Keywords: SAMOS, SOAR Telescope, Digital Micromirror Device, multi-object spectroscopy, adaptive optics, programmable slit mask, astronomical instrumentation, science commissioning

\section{INTRODUCTION}

The combination of large imaging surveys, time-domain astronomy, and increasingly ambitious spectroscopic programs has created a growing demand for highly efficient multi-object spectroscopic facilities. Conventional slit-mask spectrographs provide excellent multiplexing capabilities, but require masks to be designed and fabricated in advance, making rapid reconfiguration impractical \cite{2003SPIE.4841.1670L,2004SPIE.5492..331C}. Configurable Slit Unit (CSU) spectrographs represent an alternative approach with greater versatility, but are typically limited to a few tens of slits and configuration times of several minutes \cite{2008SPIE.7018E..0IS}. Robotic fiber-positioning systems offer rapid target allocation over large fields, but are generally best suited to relatively sparse target distributions\cite{silber_robotic_2022}. Integral-field spectrographs provide complete spatial information and are particularly effective in crowded environments, although typically over fields of only a few square arcseconds \cite{2024A&A...687A..93A,sivanandam_final_2024}.

These complementary approaches leave a gap in observational capability for programs that require both high multiplexing and rapid reconfiguration over moderately large fields of view. Digital Micromirror Devices (DMDs) offer a compelling solution by enabling programmable slit masks that can be reconfigured in real time while retaining the advantages of slit-based spectroscopy \cite{2000ASPC..195..443M,2009SPIE.7210E..0AR}. In contrast to conventional slit-mask systems, a DMD can define hundreds of independently configurable slitlets within a single field and reconfigure them in seconds, providing an attractive combination of multiplexing efficiency, operational flexibility, and rapid response to evolving scientific priorities.

The SOAR Adaptive Module Optical Spectrograph (SAMOS) was developed to address this need through the use of Digital Micromirror Device (DMD) technology \cite{2016SPIE.9908E..8VR,2016SPIE.9908E..4ZS}. Installed on the 4.1-m SOAR Telescope behind the SOAR Adaptive Module (SAM) ground-layer adaptive optics system \cite{SAM}, SAMOS combines adaptive optics-assisted imaging with programmable multi-object spectroscopy over a $3'\times3'$ field of view. Slit masks are generated electronically through DMD software control, enabling rapid reconfiguration and simultaneous operation of the spectroscopic and imaging channels.

\begin{figure}[ht]
\centering
\includegraphics[width=0.85\linewidth]{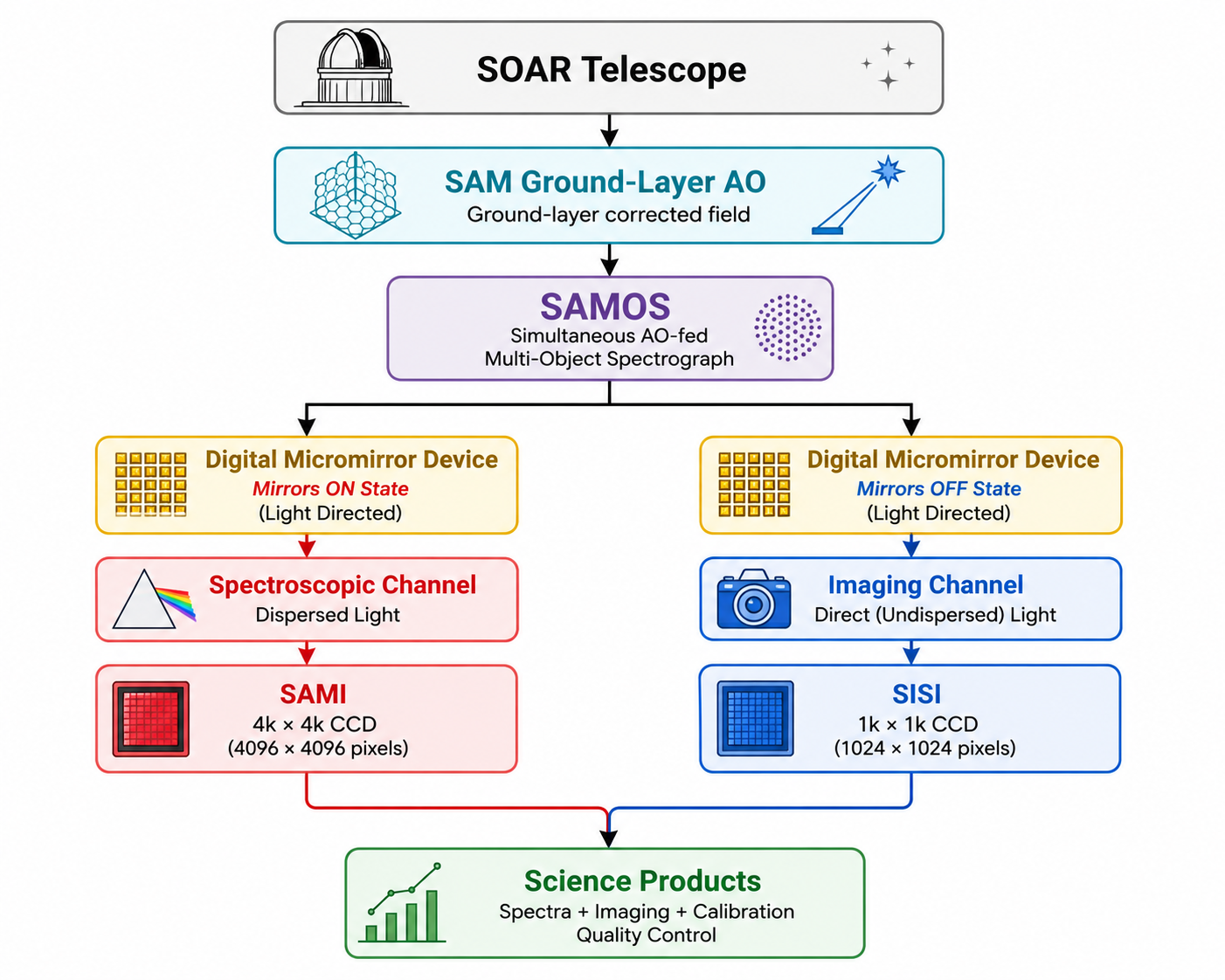}
\caption[SAMOS operational architecture]{
Simplified operational architecture of SAMOS. The AO-corrected field delivered by the SOAR Adaptive Module is projected onto a programmable Digital Micromirror Device. Individual mirrors direct light toward either the spectroscopic channel (SAMI) or the imaging channel (SISI), enabling simultaneous spectroscopy and imaging over the same field of view.
}
\label{fig:samos_architecture}
\end{figure}

The completion of science commissioning marks the transition of SAMOS from instrument development to routine scientific operations. This paper summarizes the observing procedures, operational capabilities, and on-sky performance established during commissioning. We describe the operational architecture (Figure~\ref{fig:samos_architecture}) and the end-to-end workflow, from target acquisition and programmable slit-mask generation to automated data reduction and calibration, and present representative science-verification observations that demonstrate the instrument's multiplexing capabilities and overall performance.

\section{SYSTEM OVERVIEW}

SAMOS is an adaptive-optics-assisted multi-object spectroscopic and imaging system operating at the 4.1-m SOAR Telescope. The system combines a programmable Digital Micromirror Device (DMD), a multiplexed spectroscopic channel, a parallel imaging channel, and a dedicated software environment for target acquisition, slit-mask generation, calibration, and data reduction. Together, these elements provide a unified framework for rapid and versatile spectroscopic observations over a $3'\times3'$ field of view.

At the heart of the instrument is a Texas Instruments DMD that serves as a dynamically configurable slit mask. Individual micromirrors can be addressed electronically, allowing slit patterns to be generated and modified in real time without the need for mechanically fabricated masks. This capability enables rapid reconfiguration of target lists and observing strategies, allowing efficient response to changing scientific priorities and observing conditions.

\begin{figure}[ht]
    \centering
    \includegraphics[width=0.99\textwidth]{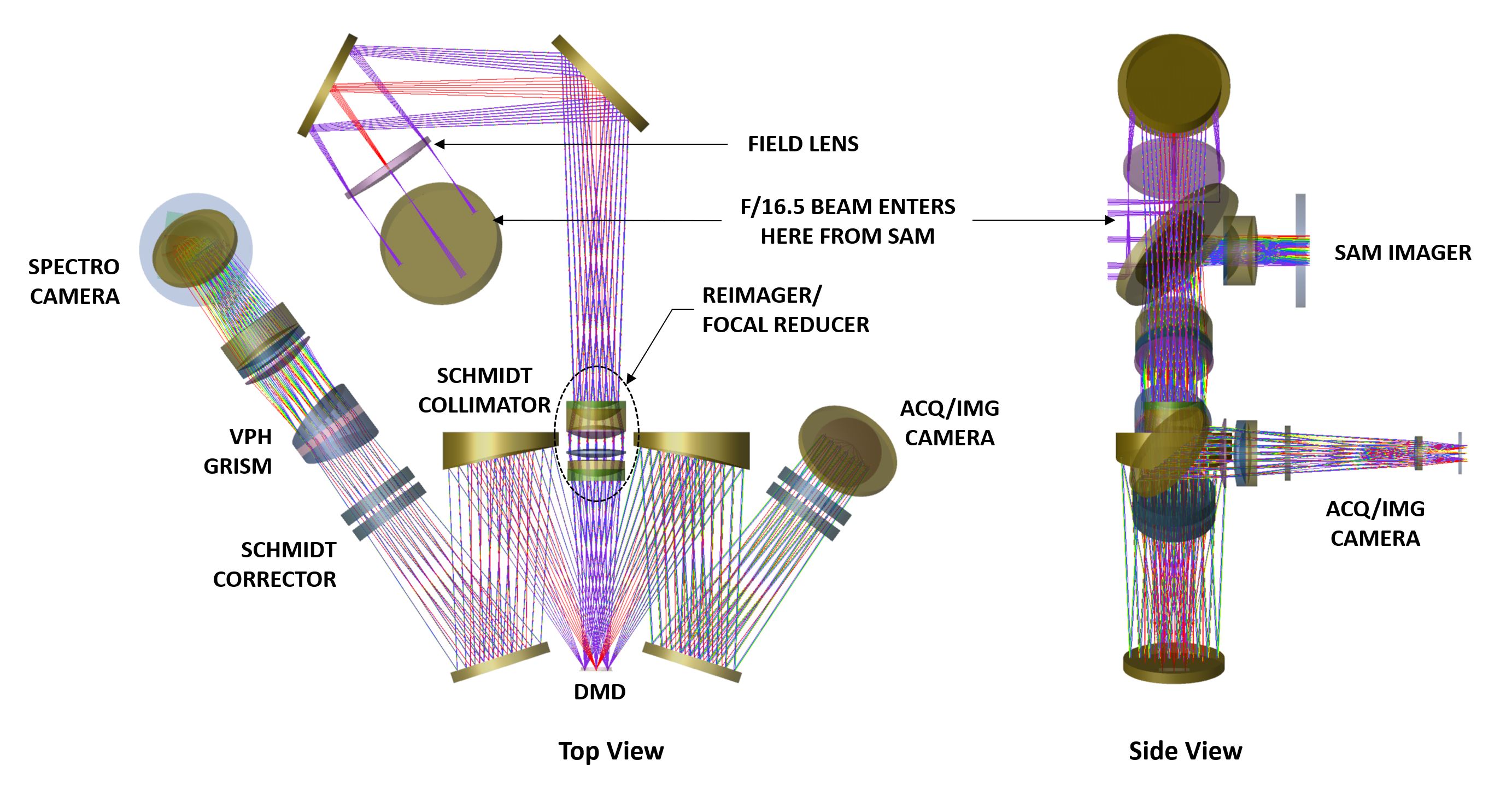}
%     \vspace{1cm}
    \caption{SAMOS optical layout, top and side view.}
    \label{fig:SAMOSLayout}
%    \vspace{1cm}
\end{figure}

\begin{figure}[ht]
    \centering
    \includegraphics[width=0.99\textwidth]{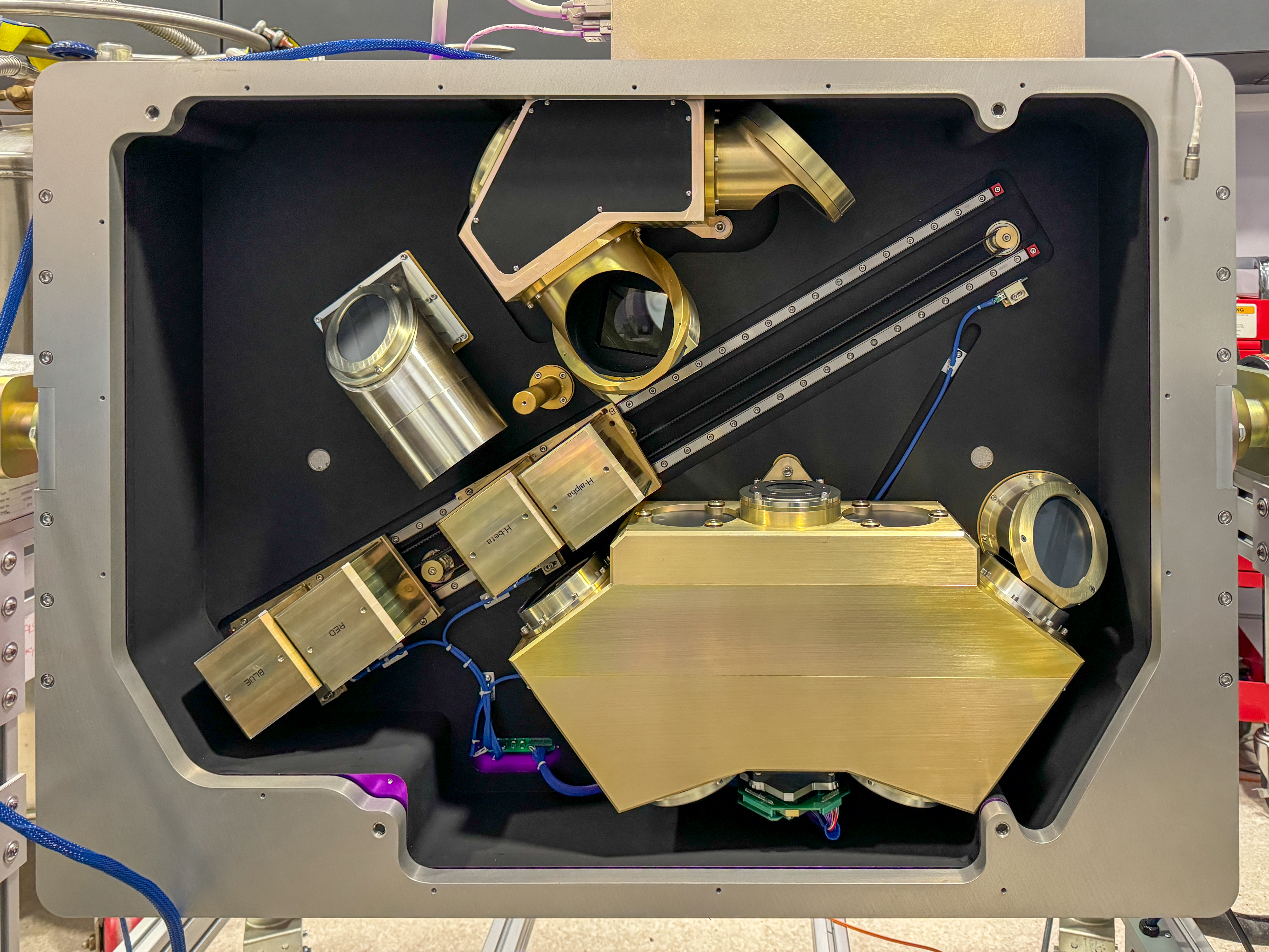}
     \vspace{5mm}
    \caption{SAMOS mechanical layout as built, oriented to match the figure to the left.}
    \label{fig:SAMOS_Photo}
    \vspace{1cm}
\end{figure}

With reference to Figure~\ref{fig:SAMOSLayout}, the AO-corrected field delivered by SAM is folded and projected onto the DMD through a reimaging system. Each DMD mirror can be addressed to reflect light toward the spectroscopic channel. The complementary mirror state feeds a parallel imaging channel. Both beams pass through identical reverse-Schmidt collimators and then either a selectable VPH grating and camera optics for the spectroscopic channel, or a filter and camera optics for the imaging channel. As a result, SAMOS can perform simultaneous spectroscopy and imaging of the same field, supporting target acquisition, astrometric registration, slit verification, and science observations within a unified observing framework. A photograph of the optomechanical layout is shown in Figure~\ref{fig:SAMOS_Photo}.

\begin{comment}
\begin{table}
\caption{Principal characteristics of SAMOS.}
\begin{tabular}{|l|l|}
\hline
Parameter & Value \\
\hline
Telescope & SOAR 4.1 m \\
Corrected field & $3'\times3'$ \\
DMD format & $2048\times1080$ mirrors \\
Mirror scale & 0.167 arcsec mirror$^{-1}$ \\
Low-resolution modes & $R\sim2500$ \\
High-resolution modes & $R\sim10,000$ \\
Spectral coverage & 4000--10000\,\AA \\
Imaging channel & SISI \\
Spectroscopic channel & SAMI \\
\hline
\end{tabular}
\end{table}
\end{comment}

%\begin{figure}
%\centering
%\begin{minipage}{.48\textwidth}
  %\centering
 % \includegraphics[width=.9\linewidth]{figures/SAMOS_Optical_Layout.JPG}
  %\captionof{figure}{SAMOS optical layout, top and side images}
  %\label{fig:OpticalLayout}
%\end{minipage}%
%\begin{minipage}{.48\textwidth}
  %\centering
%\includegraphics[width=.75\linewidth]{figures/SAMOS_AtSOAR_OptoMech.jpg}
  %\captionof{figure}{SAMOS mechanical layout as built, oriented to match the figure to the left.}
  %\label{fig:MechanicalLayout}
%\end{minipage}%
%\caption{Left: SAMOS optical layout, top and side images; right: SAMOS mechanical layout as built, oriented to match the figure to the left.}
%\label{fig:SAMOSLayout}
%\end{figure}

\subsection{Adaptive-optics-assisted operation}

SAM provides ground-layer adaptive-optics correction over the full $3'\times3'$ field of view sampled by SAMOS. Under favorable atmospheric conditions, image quality in the red optical bands can approach $0.4''$ FWHM across the corrected field.

The resulting improvement in image quality reduces source crowding, increases target contrast, and enables higher spectral resolution and lower sky background through the use of narrower slitlets without significant loss of throughput. These gains are particularly valuable for observations of crowded stellar fields and compact targets.

The SOAR telescope is equipped with an internal calibration unit with quartz and arc lamps. SAM has two dedicated guide star probes located in the telescope's uncorrected focal plane. The probes share a $5'\times5'$ patrol field to provide both tip-tilt correction and accurate telescope tracking.

\subsection{Digital micromirror slit plane}

At the core of SAMOS is a 1080$\times 2048$ Texas Instruments Cinema-2K Digital Micromirror Device, containing more than 2.2 million individually addressable mirrors. The AO-corrected field is reimaged onto the central region of the DMD, each mirror subtending approximately $0.167''$ on the sky.

Each micromirror can be independently directed toward either the spectroscopic or imaging optical channel. By selecting groups of mirrors, observers can create slitlets of arbitrary length and width, define custom slit patterns, and in general optimize target selection for any given scientific program. Because the slit configuration is entirely software controlled, reconfiguration times - of the order of a few seconds - are negligible compared with those required by conventional mask-based instruments.

The DMD architecture also supports specialized observing modes, including slitlet scanning and coded masks to implement integral-field observations through Hadamard Transform Spectroscopy.

\subsection{Spectroscopic channel}

The spectroscopic channel records the dispersed light selected by the DMD using the SAMI CCD 4K$\times$4K detector. The optical design provides simultaneous coverage of the entire $3'\times3'$ over an approximately 1350$\times$1350 pixel area, corresponding to $0.133''$/pixel, while maintaining spectral order lengths approaching 2850 detector pixels. The grating orientation is chosen such that the dispersion direction is aligned approximately with declination on the sky, with the SAMI detector mounted accordingly.

SAMOS currently supports four operational grating configurations. Two low-resolution modes provide continuous coverage of the optical spectrum from approximately 4000\,\AA\ to 10000\,\AA\ at resolving powers of $R\sim2500-3000$ for 0.33" slits (2 micromirrors). Two high-resolution modes provide resolving powers approaching $R\sim10,000$ over wavelength intervals centered on important diagnostic spectral features, including the H$\beta$+[O,III] and H$\alpha$+[S,II] regions.
A fifth grating currently under procurement will provide high-resolution spectroscopy in the Ca,II triplet region near 8500--8750\,\AA, enabling radial-velocity studies comparable to those performed by the Gaia Radial Velocity Spectrometer.

\subsection{Imaging channel}

Light reflected by DMD mirrors in the complementary state is directed toward an independent 1K$\times$1K CCD imaging channel with $0.18''$/pixel scale. The imaging camera, built by Spectral Instruments and dubbed SISI (Spectral Instruments SAMOS Imager), operates in parallel with the spectroscopic channel and provides acquisition, astrometric registration, target verification, and science imaging capabilities. 

The imaging channel supports both broadband and narrowband observations. During routine operations, imaging data are used to establish the astrometric transformation between celestial coordinates and DMD mirror positions, verify slit placement, monitor observing conditions, and provide complementary photometric information for spectroscopic targets.

The two channels operate in parallel but are fully independent. This means that, for example, during long spectroscopic exposures multiple images can be taken, in different filters, without interference between the two channels. Besides the observing efficiency gains of effectively operating two instruments in parallel, the simultaneous imaging capability represents a key operational advantage. SAMOS provides information that can be incorporated directly into the spectroscopic reduction process, including source identification, slit-loss estimation, and validation of spectral extractions.

\subsection{Operational modes}

At the conclusion of the SAMOS science commissioning campaign, the general operational architecture of SAMOS has been validated enabling the following observing modes:

\begin{enumerate}
\item Broadband and narrowband imaging;
\item Classical multi-object spectroscopy using programmable slitlet masks;
\item High-multiplex spectroscopy of crowded stellar fields through multiple point-source apertures;
\item Rapid-response acquisition of transient targets;
\item Repeated observations with dynamically updated slit configurations.
\end{enumerate}
Besides, various operational constraints like the baseline angle of the Nasmyth rotator, the procedure to rapidly acquire guide stars and the effective communication between SAMOS and the telescope control system have been validated.
The combination of adaptive-optics-assisted image quality, programmable slit selection, simultaneous imaging, and automated data reduction provides a uniquely integrated operational capability among current optical spectroscopic facilities. These characteristics were extensively exercised during commissioning and science verification observations and form the basis of the performance assessment presented in the following sections.

\section{SCIENCE OPERATIONS}

The science commissioning program was designed not only to validate the optical and spectroscopic performance of SAMOS, but also to establish a complete operational workflow from target selection to production of science-ready spectra. 
A principal objective was to demonstrate that adaptive-optics-assisted imaging, programmable slit selection, data acquisition, and calibration could be combined into an efficient workflow suitable for routine scientific observations carried out in remote observing mode.

\begin{figure}[ht]
\centering
\includegraphics[trim={20cm 0 20cm 0},width=0.8\linewidth]{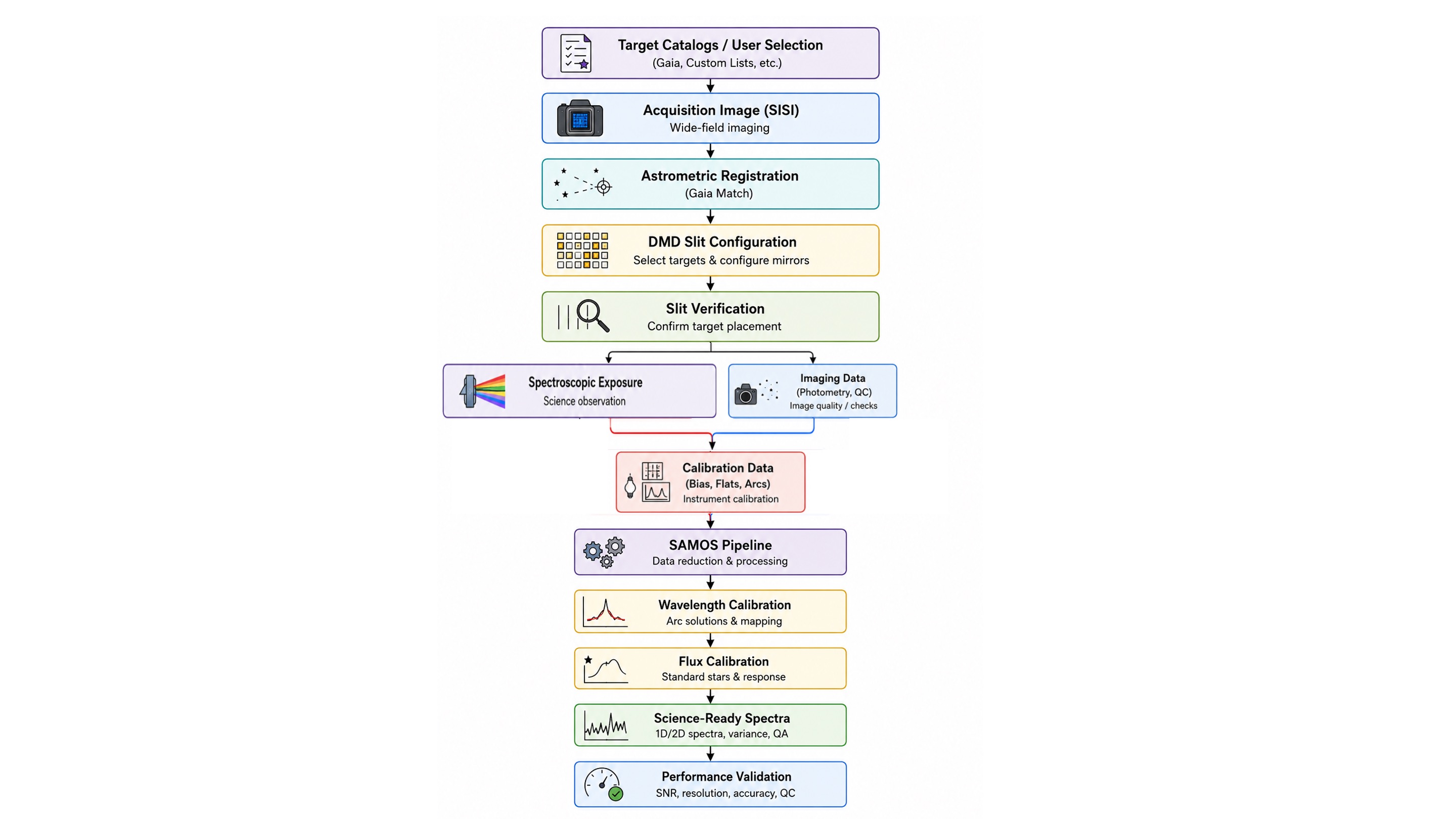}
\caption[SAMOS operational workflow]{
Operational workflow developed during SAMOS science commissioning. Acquisition images are used for astrometric registration and DMD mask generation. Spectroscopic and imaging observations are combined with calibration data and processed through the automated reduction pipeline to produce science-ready spectra.
}
\label{fig:samos_workflow}
\end{figure}

Figure~\ref{fig:samos_workflow} summarizes the operational sequence currently adopted during SAMOS observations.

\subsection{Target catalog preparation}

The preparation of a SAMOS observation begins with the definition of a target catalog for the field of interest. For each target, the catalog specifies a unique identifier, celestial coordinates (Right Ascension and Declination), the width and length of the associated slitlet, and the slit position angle. These parameters define the geometry of the programmable slit mask that will be projected onto the Digital Micromirror Device (DMD).

A software tool is used to verify the proposed slit configuration and identify potential spectral overlaps on the detector. This verification step is particularly important in crowded fields, where the high multiplexing capability of SAMOS can lead to conflicting slit assignments if not properly constrained.

The catalog is stored in the form of an ASCII DS9 region file (\texttt{.reg}), allowing the slit geometry to be displayed directly on astronomical images using SAOImage DS9. This representation provides a convenient visual verification of target selection and slit placement prior to the observing run. Multiple catalogs corresponding to different scientific priorities or observing conditions can be prepared in advance and loaded during an observing session, enabling rapid reconfiguration of the DMD slit pattern. In addition, the Instrument Control Software provides a GINGA-based interface that allows last-minute adjustments to individual slitlets or global slit-mask parameters without requiring regeneration of the entire catalog.

\subsection{Acquisition image  and astrometric registration}

Observations begin with acquisition images obtained through the SISI imaging channel. These images are processed automatically to identify stellar sources and derive an astrometric solution through cross-matching with external catalogs, primarily Gaia. The resulting transformation establishes the correspondence between celestial coordinates and detector pixels.

A daytime calibration procedure determines the reference position of the guide probe relative to the imaging field. Once the astrometric solution has been established, this calibration is used to compute the offset required to acquire a suitable guide star within the adjacent patrol field. The procedure enables rapid target acquisition and minimizes setup overheads during observing operations.

Because the SOAR internal observatory network is isolated from external internet access, all astrometric and photometric catalogs required for an observing program must be uploaded to the SAMOS workstation prior to the observing run. This ensures that astrometric registration, target selection, and slit-mask generation can proceed without dependence on external network resources. Another important operational constraint is that for observations conducted with GLAO correction, laser propagation requests must be submitted several days in advance in accordance with observatory procedures.

\subsection{Slit-mask generation}

\begin{figure}
\centering
\begin{minipage}{.33\textwidth}
  \centering
  \includegraphics[width=.9\linewidth]{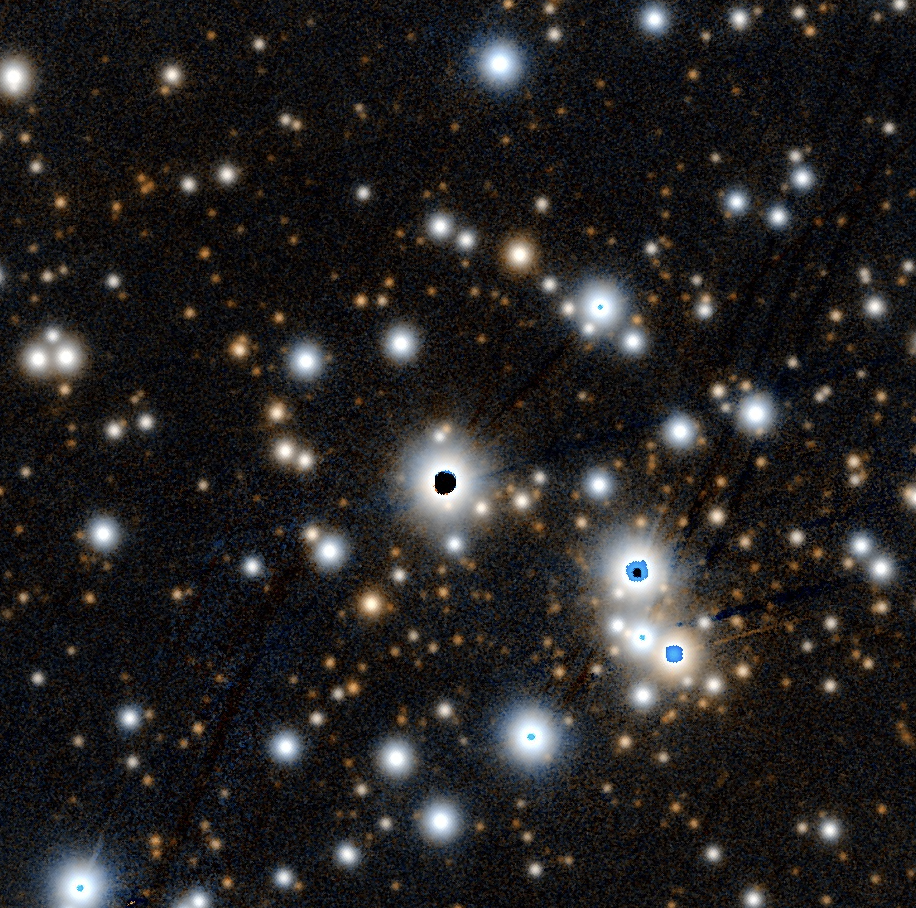}
  %\captionof{figure}{Target Field, part of\\ the Dolidze~25 region, \\from PanSTARRS DR1.}
  %\label{fig:test1}
\end{minipage}%
\begin{minipage}{.33\textwidth}
  \centering
  \includegraphics[width=.95\linewidth]{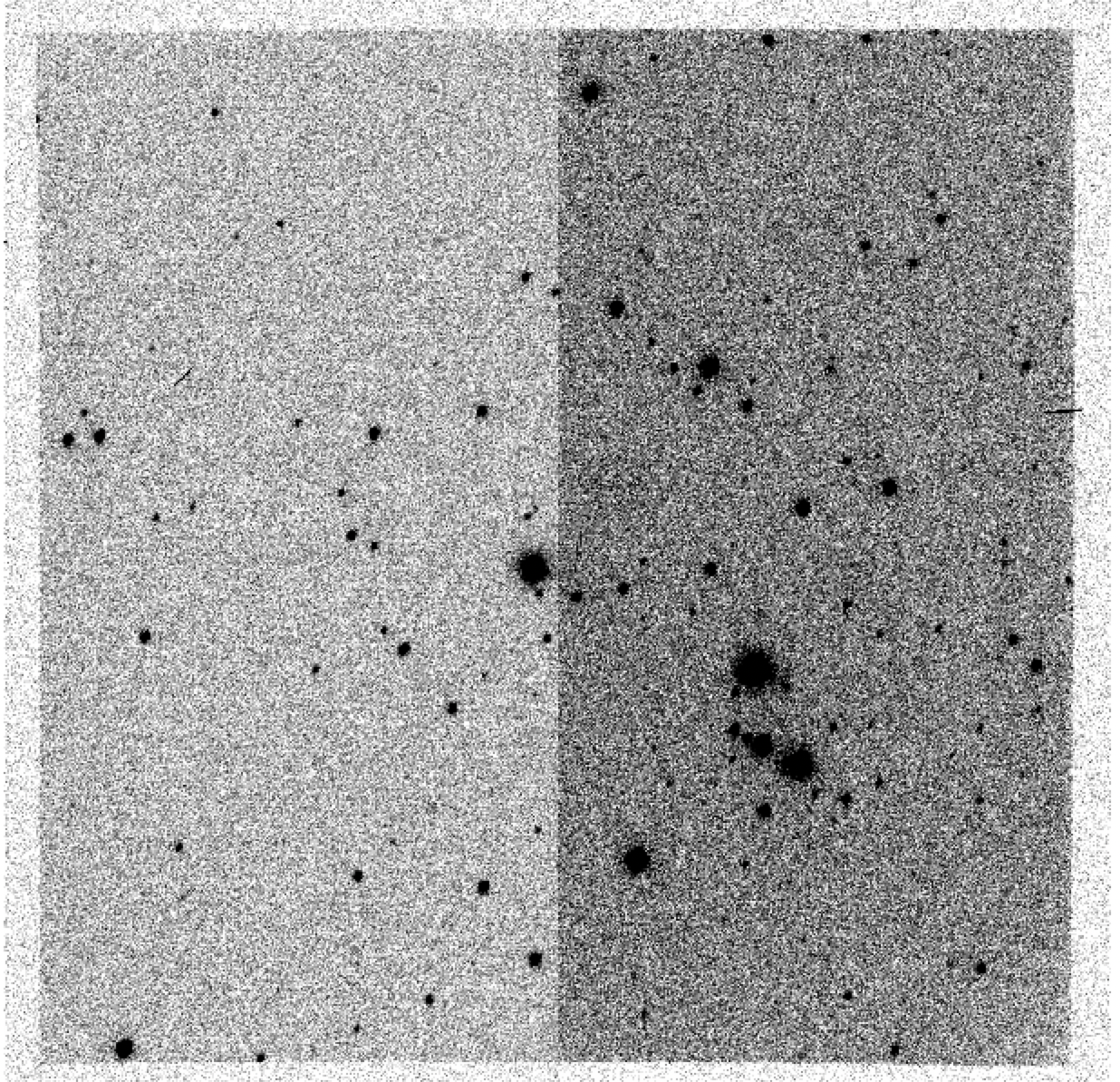}
  %\captionof{figure}{Target acquisition image, i-band, 5~s exposure}
  %\label{fig:test1}
\end{minipage}%
\begin{minipage}{.33\textwidth}
  \centering
  \includegraphics[width=.95\linewidth]{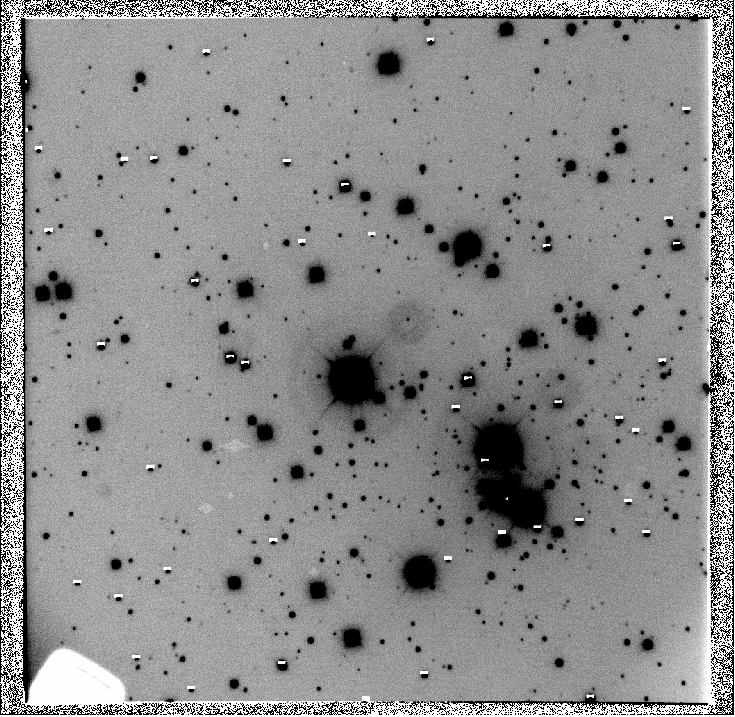}
  %\captionof{figure}{Science exposure with slits, i-band, 100 s.. }
  %\label{fig:test2}
\end{minipage}
\caption[Dolidze 25 acquisition and slit verification images]{Left: target field in the Dolidze~25 region from the Pan-STARRS DR1 survey. Center: target-acquisition image obtained with the SISI imaging channel through the $i$ filter (5~s exposure), shown without processing. Right: imaging exposure obtained through the same filter with the DMD slit-mask configuration projected onto the field (100~s exposure), after flat-field correction. The guide-star probe arm is visible entering the field from the lower-left corner.}
\label{fig:samos_imaging}
\end{figure}

Once the astrometric solution has been established, the target catalog is transformed into a DMD slit configuration for spectroscopic observations (Fig.~\ref{fig:samos_imaging}). Target coordinates and associated slit parameters are imported from the user catalog and converted from celestial coordinates to detector pixels and subsequently to DMD coordinates, producing the slit pattern to be projected onto the micromirror array.

The mapping between the sky, detector, and DMD focal planes is established through a set of calibrated coordinate transformations. The transformation between DMD coordinates and detector pixels is derived from calibration images obtained using a synthetic pinhole grid projected by the DMD. The transformation between celestial coordinates and detector pixels is determined for each pointing from the acquisition images obtained with the SISI camera and matched to Gaia astrometric reference catalogs. Together, these calibrations define the mapping between sky coordinates and DMD mirrors required for accurate slit placement.

The slit configuration can be verified directly on the sky using the imaging channel prior to the start of the spectroscopic sequence. An optimization routine may further refine slit positions using source centroids measured from the acquisition image. The geometry of each slitlet can also be adjusted to match the scientific objectives of the observation. Narrow slitlets may be selected when image quality is excellent and maximum spectral resolution is desired, whereas wider slitlets can be used to maximize throughput.

The ability to validate and modify slit configurations immediately before an observation provides substantial operational freedom. Updated target lists, revised scientific priorities, and changing observing conditions can be accommodated without the need to manufacture or exchange physical masks. In routine operations, a pre-defined slit configuration can typically be verified, optimized, and uploaded to the DMD in less than one minute, while the actual upload process requires only a few seconds.
% Side-by-side figures

\subsection{Spectroscopic observations}

\begin{figure}[t]
\centering
\includegraphics[trim={0cm 0cm 0cm 0cm},width=0.8\linewidth]{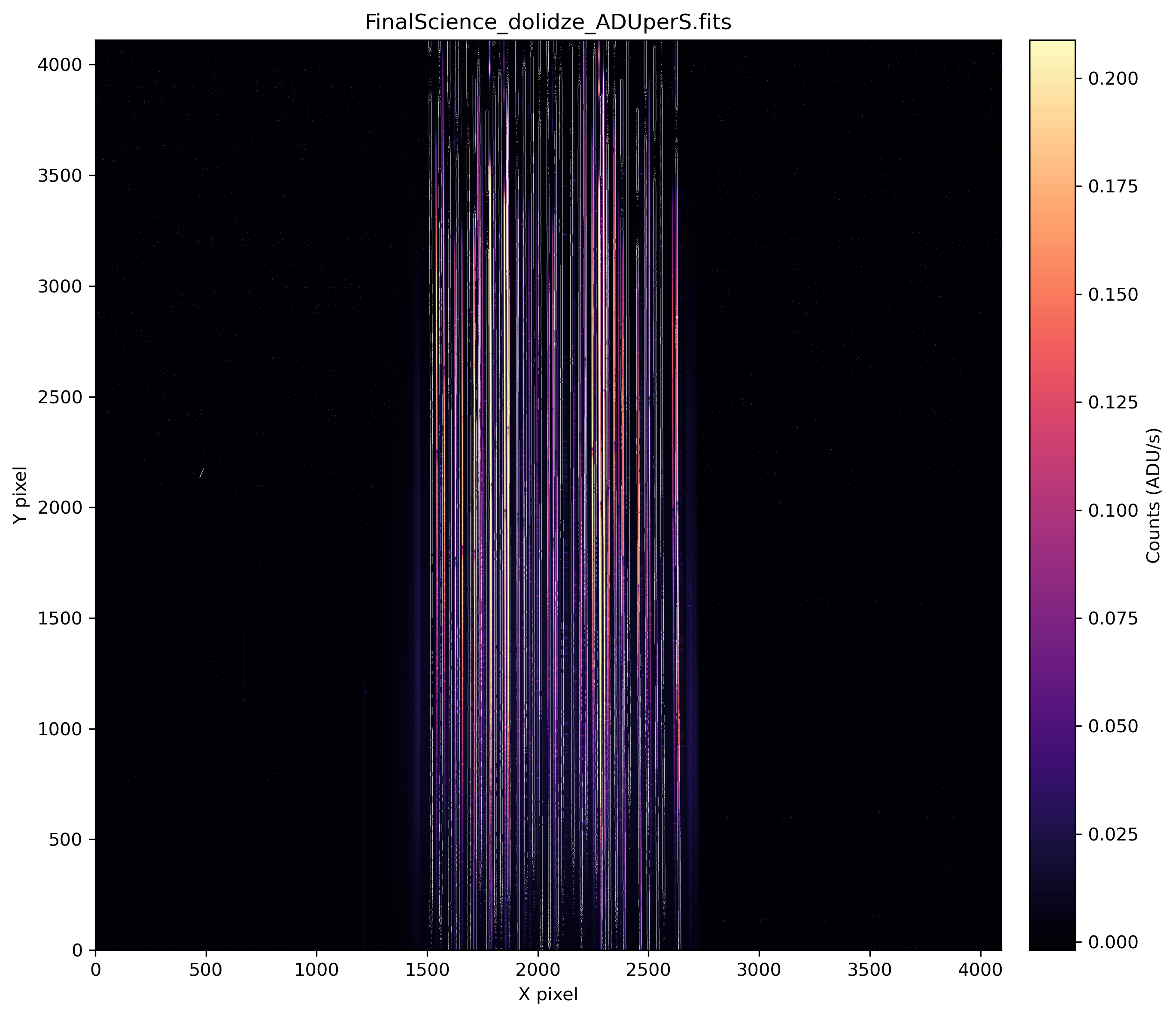}
\caption[SAMOS multiplexed spectra]{Multiplexed spectra obtained with SAMOS in the Dolidze~25 field using the low-resolution red grating (600--950\,nm wavelength range). The image is the average of three 900~s science exposures.}
\label{fig:spectra}
\end{figure}

Once the slit configuration has been verified, spectroscopic exposures are acquired using the selected grating configuration (Fig.~\ref{fig:spectra}). Spectroscopic data are recorded using the facility SAMI detector, which is presently operated through its native control software independently of the SAMOS control environment. Since SAMI was originally designed as the imaging camera for the SAM adaptive-optics module, dedicated quick-look tools for multi-object spectroscopy are currently limited. Likewise, the native FITS headers do not yet contain the complete set of SAMOS configuration parameters, which are incorporated during subsequent data processing.

When required, the spectroscopic channel can also be operated in direct-imaging mode by removing the dispersing element and configuring the DMD with all mirrors directed toward the spectroscopic arm. These white-light images provide an independent view of the field at the SAMI focal plane and can be used for source identification, astrometric verification, and validation of the spectroscopic observations.

%Because slit configurations can be regenerated rapidly, multiple target selections can be observed within a single observing session without the overheads associated with physical mask exchanges.

\subsection{Simultaneous imaging support}

A distinctive operational feature of SAMOS is the simultaneous acquisition of imaging data of the same field during spectroscopic observations. These images support several aspects of instrument operation and data calibration.

First, they provide direct verification of target acquisition and slit placement. In addition, automated analysis routines measure image-quality metrics, including the median full width at half maximum (FWHM), and estimate the photometric zeropoint through comparison with all-sky photometric catalogs, using SkyMapper in the southern hemisphere and Pan-STARRS in the northern hemisphere. These measurements provide a continuous assessment of observing conditions and data quality.

Imaging observations acquired in the $griz$ bands also support the final stages of spectrophotometric calibration. Broadband photometry obtained from the imaging channel can be compared with synthetic photometry derived from the extracted spectra, providing an independent constraint on the absolute flux calibration.

Finally, the imaging data facilitate the interpretation of observations in crowded fields by providing information on source morphology, local crowding conditions, and possible contamination from neighboring objects. This capability proved particularly valuable during science-verification observations of stellar clusters, where accurate slit placement and source identification are critical to the success of multiplexed spectroscopic observations.

\subsection{Calibration observations}

In addition to science exposures, SAMOS observations require the acquisition of calibration data, including bias frames, quartz-lamp flats, and arc-lamp exposures. Because the DMD slit pattern defines part of the optical configuration of the instrument, calibration observations must be obtained for each slit-mask configuration used for the corresponding science exposures.

Quartz-lamp flats are used to determine slit traces, pixel-to-pixel sensitivity variations, and illumination corrections, while arc-lamp exposures provide the wavelength calibration. Calibration sequences are typically acquired at the end of each observing night, although additional calibrations may be obtained whenever slit configurations are modified during the observing run.

Science commissioning demonstrated that the complete calibration workflow can be executed routinely and reproducibly. The successful integration of calibration acquisition, automated reduction procedures, and quality-control diagnostics established the baseline operational model now adopted for routine SAMOS observations.

\subsection{SAMOS pipelines}

At the completion of an observing sequence, all science and calibration data are processed through the SAMOS reduction environment, which includes dedicated pipelines for spectroscopic and imaging observations.

The spectroscopic pipeline performs detector calibration, slit-trace determination, wavelength calibration, spectral extraction, atmospheric correction, and flux calibration through a largely automated workflow. In parallel, the imaging pipeline performs detector calibration, flat-field correction, source extraction, and photometric calibration using all-sky reference catalogs. For broad-band observations, photometric zeropoints are derived through comparison with SkyMapper or Pan-STARRS catalog photometry.

The resulting products include wavelength-calibrated and flux-calibrated one-dimensional spectra, calibrated imaging data, quality-control diagnostics, and intermediate calibration products. This automated environment minimizes observer intervention and enables rapid assessment of data quality during observing runs.

A final processing stage combines the spectroscopic and imaging products. Synthetic photometry derived from the spectra is compared with the calibrated broad-band imaging measurements, and the resulting differences are used to refine the spectral response function. This procedure places the spectra on an absolute flux scale and produces science-ready data products in physical units.

The reduction procedures established during science commissioning have been adopted as the baseline processing workflow for routine SAMOS observations.

\section{COMMISSIONING PROGRAM}

The science commissioning campaign was designed to validate the operational readiness of SAMOS and to establish a complete observing workflow for routine scientific use. The program exercised all major instrument functions, including target catalog preparation, field acquisition, astrometric registration, slit-mask generation, low- and high-resolution spectroscopy, calibration observations, and automated data reduction. In parallel, observatory-level interfaces such as guide-star acquisition, laser operation, and telescope-control-system interactions were tested and refined. Together, these activities provided an end-to-end validation of the instrument, from target selection through the delivery of science-ready data products.

Particular attention was devoted to the evaluation of multiplexed spectroscopy in crowded stellar fields, one of the principal science drivers for the instrument. These observations tested the accuracy of the astrometric registration, the reliability of slit placement, and the robustness of the extraction and calibration procedures under realistic observing conditions.

The open cluster Dolidze~25 was selected as the primary science-verification target. Its rich stellar population, broad dynamic range in brightness, and compact angular extent provide an ideal testbed for evaluating multiplexing efficiency, extraction robustness, wavelength-calibration accuracy, and photometric consistency across a large number of simultaneously observed sources. The observations discussed below demonstrate the complete operational chain from target acquisition to calibrated science spectra and form the basis of the performance assessment presented in the following section.

\section{PERFORMANCE RESULTS}

\subsection{Astrometric performance}

Accurate registration between celestial coordinates and DMD mirror positions is essential for efficient operation of SAMOS. The transformation between the DMD and the SISI detector was calibrated using a synthetic grid of $11\times11$ point sources projected onto the micromirror plane. Repeated measurements demonstrated that this mapping is stable to well below one detector pixel and does not require frequent recalibration.

Astrometric solutions derived from acquisition images revealed measurable field distortion across the full field of view. To account for these effects, the World Coordinate System (WCS) solution incorporates higher-order Simple Imaging Polynomial (SIP) coefficients. Because not all science fields contain a sufficient number of Gaia reference stars to constrain the distortion model directly, pre-calibrated SIP coefficients can be loaded from reference calibration files and combined with the astrometric solution derived from the acquisition image.

The resulting transformations provide reliable mapping between sky coordinates and DMD mirror positions over the full field of view. In practice, the dominant limitation to slit-placement accuracy is not the astrometric solution itself, but the finite size of the DMD mirrors, corresponding to $0.17''$ per mirror on the sky. When higher positioning accuracy is required, slit locations can be refined through centroid measurements obtained from the acquisition image. In such cases, preservation of the final DMD configuration is essential to ensure consistency between the science observations and the associated calibration data.

The combination of simultaneous imaging and programmable slit selection proved particularly effective in crowded stellar fields, where source identification and accurate slit placement are critical. Repeated observations confirmed both the opto-mechanical stability of the instrument and the repeatability of the acquisition procedure, demonstrating the suitability of the operational workflow for routine science observations.

\subsection{Spectroscopic performance}

During science commissioning, observations were obtained in both low-resolution and high-resolution modes, exercising the full range of spectroscopic capabilities currently available with SAMOS.

\begin{figure}[t]
\centering
\includegraphics[trim={0cm 0cm 30cm 14cm}, clip, width=1.0\linewidth]{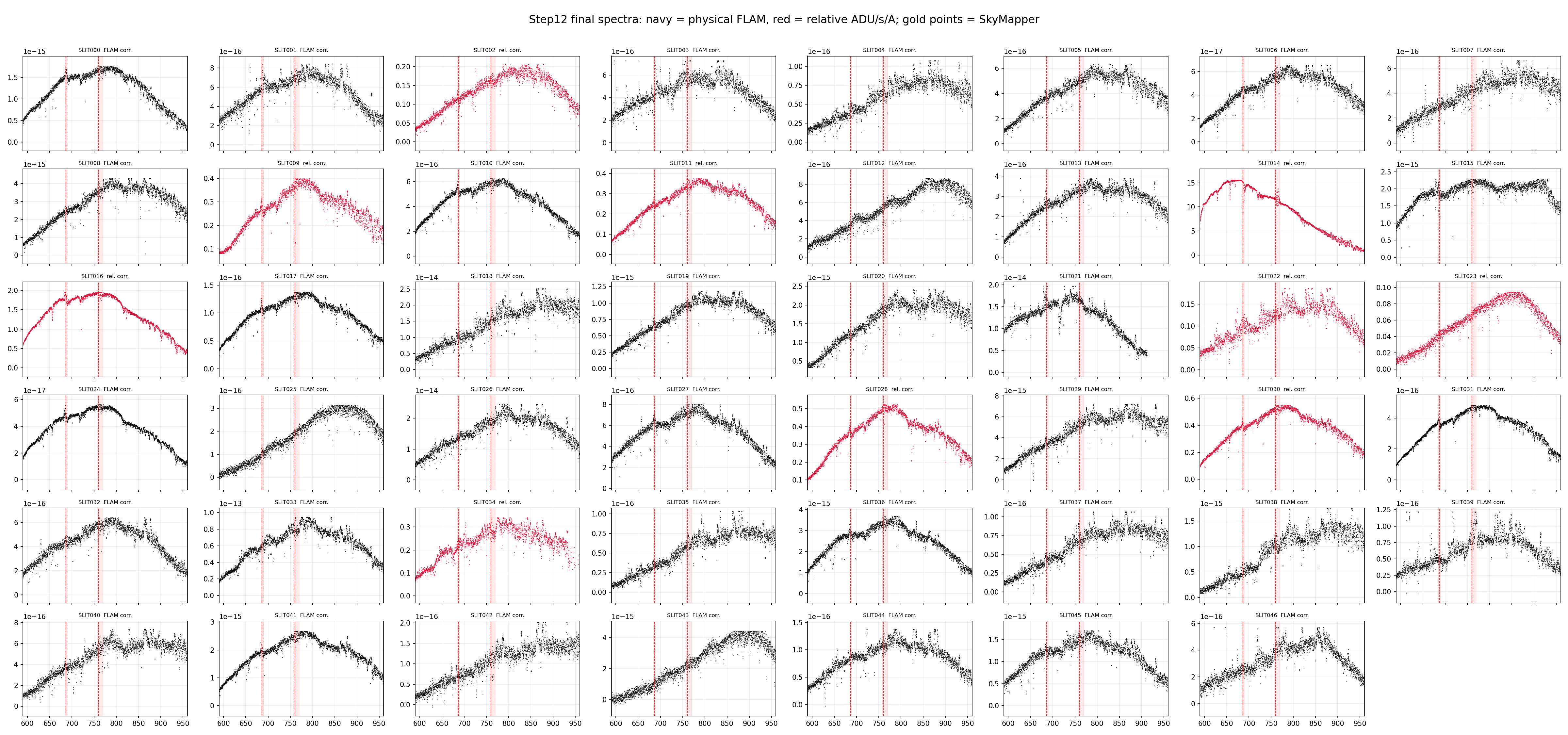}
\caption[Multiplexed extracted spectra]{
Mosaic of spectra extracted from a single SAMOS observation of the Dolidze~25 field. Each panel corresponds to an individual slitlet and spans the wavelength range covered by the low-resolution red grism. Black spectra are shown on an absolute flux scale, while red spectra correspond to sources that could not be directly cross-calibrated against the SkyMapper catalog and are therefore displayed in relative flux units.  The diversity of spectral shapes reflects the wide range of source brightnesses and colors present in the field. This figure illustrates the multiplexing capability of SAMOS and the ability of the reduction pipeline to process a large number of spectra simultaneously.
}
\label{fig:multiplex_spectra}
\end{figure}

The spectroscopic commissioning program evaluated the wavelength coverage, spectral resolution, calibration stability, and extraction performance of the instrument. Arc-lamp exposures obtained through the programmable slit masks were used to derive wavelength solutions for all slitlets simultaneously, while quartz-lamp exposures were used to identify the individual spectral traces and correct for pixel-to-pixel sensitivity variations.

The global wavelength solution derived from the arc-lamp exposures achieved residuals of approximately 0.3\,\AA. Subsequent refinement using atmospheric OH emission features further reduced slit-to-slit systematic offsets and ensured a consistent wavelength scale across the full multiplexed dataset.

The extracted spectra were subsequently combined with broad-band photometry obtained from acquisition images and from imaging observations associated with different slit-mask configurations. Since a target observed spectroscopically in one configuration is often unobscured in another, the imaging data collected throughout the observing sequence provide photometric measurements for a large fraction of the spectroscopic sample. Together with SkyMapper catalog photometry for bright saturated sources, these measurements were used to place the spectra on an absolute flux scale, producing the final science-ready spectra shown in Figure~\ref{fig:multiplex_spectra}.

The complete reduction sequence was tested end-to-end using observations obtained with the low-resolution red grism covering the nominal 6000--10000\,\AA\ wavelength range. The successful execution of the workflow, from detector-level calibrations through wavelength and flux calibration, demonstrated the operational readiness of the SAMOS spectroscopic reduction environment for routine scientific observations.

\subsection{Multiplexing capability}

One of the principal advantages of SAMOS is its ability to observe large numbers of targets simultaneously through electronically configurable slit masks. During the Dolidze~25 commissioning observations two sets of exposures were taken with 62 and 60 slitlets 
simultaneously deployed. Each slitlet was 9 mirrors wide, corresponding to about 11 SAMI pixels.  For faint sources, this setup allowed determining a local sky spectrum to subtract using an optimal extraction algorithm for each source. 
\begin{figure}[t]
\centering
\includegraphics[width=0.4\linewidth]{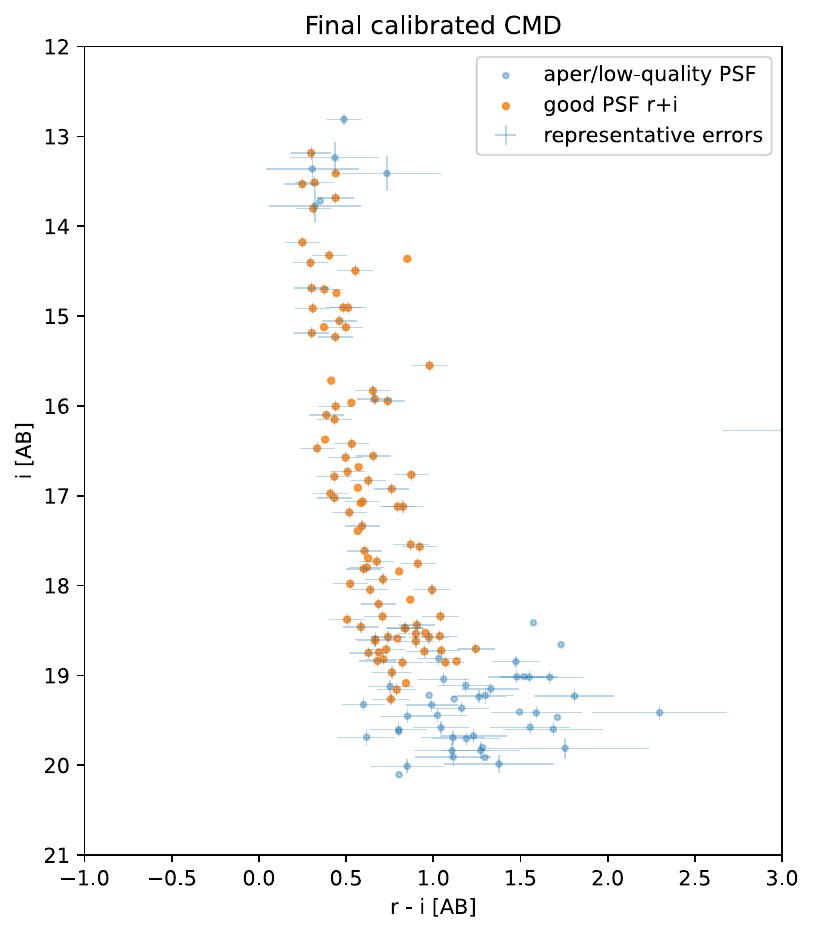}
\caption[Dolidze 25 calibrated color-magnitude diagram]{
Calibrated color--magnitude diagram of the Dolidze~25 commissioning field derived from SISI imaging observations. Sources passing the PSF-quality selection criteria are shown in orange, while lower-quality measurements are shown in blue. Representative photometric uncertainties are indicated. The well-defined stellar sequence demonstrates the quality of the astrometric registration, photometric calibration, and source-extraction procedures implemented in the SAMOS data-reduction pipeline.
}
\label{fig:dolidze25_cmd}
\end{figure}

%Figure~\ref{fig:dolidze25_demo} illustrates the complete observational sequence. The upper panels show the acquisition image and the corresponding DMD slit-mask configuration, while the lower panels present the resulting detector frame and representative extracted spectra. The figure highlights the ability of SAMOS to transform a crowded stellar field into a large set of independently calibrated spectra through a fully automated workflow.

To assess the possibility of packaging a larger number of spectra, reducing the width of the slitlets in the cross-dispersion direction, the same data were processed developing a procedure that determines and subtracts the sky spectrum performing a global analysis of all stellar spectra collected during an exposure. The details of this procedure will be presented in a dedicated paper (Robberto et al, to be submitted).
The programmable nature of the DMD enables rapid reconfiguration of target lists without the need for mechanical mask exchanges. This capability significantly reduces observing overheads and increases observing efficiency for spectroscopic surveys, transient follow-up observations, and studies of crowded stellar populations.

\subsection{Photometric performance}

The imaging channel was used extensively during commissioning to assess the photometric performance of SAMOS and to provide an independent calibration reference for the spectroscopic observations. Final photometric catalogs were produced using PSF-fitting photometry calibrated against the SkyMapper survey in the standard $riz$ photometric system.

For the Dolidze~25 commissioning field, the final calibrated catalog contains more than 150 sources in each of the $r$, $i$, and $z$ bands. Photometric zero points were derived from stars cross-matched with SkyMapper, yielding calibration uncertainties of 0.015, 0.012, and 0.009 mag in the $r$, $i$, and $z$ bands, respectively. The corresponding residual scatters are approximately 0.07 mag in all three filters, indicating a stable and internally consistent calibration over the full field of view. 

Figure~\ref{fig:dolidze25_cmd} presents a final calibrated color--magnitude diagram . The main stellar sequence is clearly recovered, demonstrating both the quality of the imaging data and the effectiveness of the calibration procedure. Analysis of the photometric residuals shows no significant dependence on detector position, indicating that flat-fielding, astrometric registration, and photometric calibration are stable across the imaging field. 

These results demonstrate that the SISI imaging channel provides reliable photometric measurements suitable not only for target acquisition and observing-quality assessment, but also for the final spectrophotometric calibration of SAMOS spectra.
\begin{figure}[t]
\centering
\includegraphics[width=1.0\linewidth]{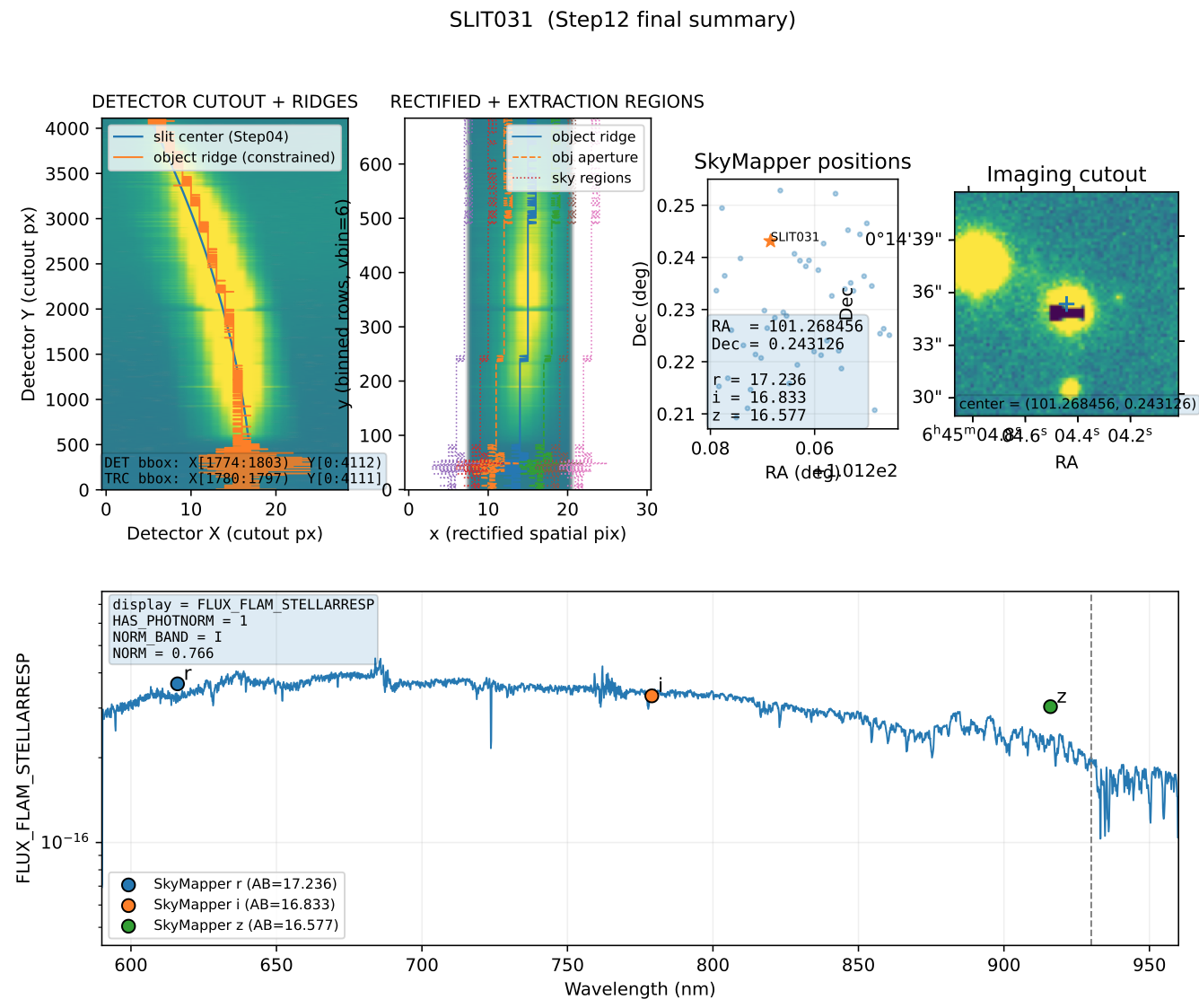}
\caption[End-to-end processing summary]{
End-to-end processing summary for a representative target observed during the Dolidze~25 commissioning program. Upper left: detector cutout showing the traced spectrum on the SAMI detector. Upper center: rectified slitlet with the object and sky extraction regions adopted by the pipeline. Upper center-right: association of the extracted spectrum with the corresponding SkyMapper catalog source. Upper right: SISI imaging-channel cutout showing the target position and slit placement. Bottom: final flux-calibrated spectrum with photometric reference measurements in units of FLAM [erg cm$^{-2}$ s$^{-1}$ Angstrom$^{-1}$]. The figure illustrates the complete data lineage from detector-level observations to the final calibrated spectroscopic product.
}
\label{fig:final_spctrophotometry}
\end{figure}

\subsection{End-to-end calibration products}

One of the distinctive capabilities of SAMOS is the integration of imaging and spectroscopy within a common operational framework. Information derived from the imaging channel, including astrometric registration, source identification, and calibrated photometry, is incorporated directly into the spectroscopic reduction process.

Figure~\ref{fig:final_spctrophotometry} presents a representative end-to-end processing summary generated by the SAMOS pipeline. The figure links the original detector data, spectral extraction geometry, catalog cross-identification, imaging counterpart, and final calibrated spectrum within a single diagnostic product. This capability provides complete provenance for each extracted spectrum and allows the observer to validate every stage of the reduction process.

The successful generation of these products demonstrates the integration of instrument hardware, observing procedures, calibration data, and automated reduction software into a unified environment capable of producing science-ready spectra from multiplexed observations.

Figure~\ref{fig:dolidze25_cmd} presents the calibrated
color--magnitude diagram derived from the Dolidze~25
commissioning observations. The recovered stellar sequence
demonstrates the quality of the astrometric registration,
PSF photometry, and photometric calibration procedures.
The resulting catalog contains more than 150 calibrated
sources with typical photometric uncertainties below
0.1 mag in the brightest magnitude range.

\section{SUMMARY AND FUTURE DEVELOPMENTS}

The completion of science commissioning marks the transition of SAMOS from an instrument-development project to a fully operational facility instrument at the SOAR Telescope. The commissioning campaign validated the complete end-to-end observing environment, from adaptive-optics correction delivered by the SOAR Adaptive Module (SAM), through programmable slit-mask generation and multiplexed spectroscopy, to automated reduction and calibration of the resulting data products. Together, these activities established a robust operational framework for routine scientific observations. The principal performance metrics achieved during commissioning are summarized in Table~\ref{tab:summary}.
\begin{table}[t]
\caption[SAMOS commissioning performance metrics]{Summary of SAMOS commissioning performance metrics.}
\label{tab:performance}
\centering
\begin{tabular}{|l|c|}
\hline
Parameter & Performance \\
\hline
Field of view & $3'\times3'$ \\
DMD plate scale & 0.167 arcsec mirror$^{-1}$ \\
Low-resolution wavelength coverage & 4000--10000\AA \\
High-resolution resolving power & $\sim10,000$ \\
Multiplex demonstrated & 62 slitlets \\
Wavelength calibration RMS & $\sim0.03$ nm \\
Spectral length & $\sim3000$ pixels \\
SkyMapper calibrators matched & 78 \\
Calibrated field sources & 213 \\
r-band ZP uncertainty & 0.015 mag \\
i-band ZP uncertainty & 0.012 mag \\
z-band ZP uncertainty & 0.009 mag \\
Photometric calibration scatter & $\sim$0.07 mag \\
Simultaneous imaging & Yes \\
Slit-mask reconfiguration time & Seconds \\
Automated reduction pipeline & Operational \\
\hline
\end{tabular}
\label{tab:summary}
\end{table}

SAMOS demonstrates the successful integration of adaptive-optics-assisted imaging, Digital Micromirror Device technology, simultaneous imaging and spectroscopy, and automated data reduction within a single observational platform. The ability to generate and reconfigure slit masks within seconds, acquire and calibrate large numbers of spectra simultaneously, and combine spectroscopic and photometric information within a unified workflow provides a powerful new capability for observations of crowded stellar fields, spectroscopic surveys, and rapid-response programs. The programmable nature of the DMD architecture also offers a path toward substantially higher multiplexing, potentially enabling observations of several hundred targets within a single field.

Future developments will include deployment of a high-resolution grating covering the Ca\,II triplet region, improved integration with the telescope software infrastructure, further automation of observing procedures, and continued refinement of the reduction and calibration pipelines. These enhancements will expand the scientific reach of the instrument while building on the operational foundation established during science commissioning.

\acknowledgments % equivalent to \section*{ACKNOWLEDGMENTS}       
 
This work is based on observations obtained with the Southern Astrophysical Research (SOAR) Telescope. We gratefully acknowledge the support of the SOAR staff throughout the commissioning and operation of SAMOS, in particular J. Elias and C. Brice\~no for their invaluable assistance.

The development of SAMOS was supported by the National Science Foundation under grant AST-1611276, with additional support provided through the Director's Discretionary Research Fund of the Space Telescope Science Institute.

% References
\bibliography{report} % bibliography data in report.bib
\bibliographystyle{spiebib} % makes bibtex use spiebib.bst

\end{document}